\documentclass[runningheads]{llncs}
\usepackage{graphicx}
\usepackage{comment}
\usepackage{amsmath,amssymb} 
\usepackage{color}

\usepackage{bbm}
\usepackage{bm}

\usepackage[utf8]{inputenc} 
\usepackage{url}            
\usepackage{booktabs}       
\usepackage{amsfonts}       
\usepackage{nicefrac}       
\usepackage{microtype}      
\usepackage{color}
\usepackage[labelformat=empty]{subfig}
\usepackage{booktabs}
\usepackage{array}
\usepackage{multirow}
\usepackage{mathtools}
\usepackage{stmaryrd}
\usepackage[para,online,flushleft]{threeparttable}
\usepackage{float}
\usepackage{comment}
\usepackage{amsmath}
\usepackage{svg}
\usepackage{xcolor}

\newcolumntype{P}[1]{>{\centering\arraybackslash}p{#1}}

\usepackage[linesnumbered,ruled,vlined]{algorithm2e}

\SetCommentSty{mycommfont}
\SetKwInput{KwInput}{Input}                
\SetKwInput{KwOutput}{Output}

\makeatletter
\DeclareRobustCommand\onedot{\futurelet\@let@token\@onedot}
\def\@onedot{\ifx\@let@token.\else.\null\fi\xspace}

\makeatother

\newcommand*\samethanks[1][\value{footnote}]{\footnotemark[#1]}

\makeatletter
\renewcommand{\paragraph}{%
  \@startsection{paragraph}{4}%
  {\z@}{0.5\baselineskip \@plus 0ex \@minus 0ex}{-1em}%
  {\normalfont\normalsize\bfseries}%
}
\makeatother

\begin{document}

\pagestyle{headings}
\mainmatter
\title{Social Distancing 2.0 with Privacy-Preserving Contact Tracing to Avoid a Second Wave of COVID-19} 

%
\titlerunning{Social Distancing 2.0 with Privacy-Preserving Contact Tracing}
%
\author{Yu-Chen Ho\thanks{Both authors contributed equally to this work.}\inst{1} \and
        Yi-Hsuan Chen\samethanks\inst{2}\and
        Shen-Hua Hung\inst{1} \and \\
        Chien-Hao Huang\inst{1} \and 
        Poga Po\inst{1}\and 
        Chung-Hsi Chan\inst{1} \and 
        Di-Kai Yang\inst{1} \and \\
        Yi-Chin Tu\thanks{Corresponding authors. {Email: {\tt  contact@taimedimg.tw, fangct@ntu.edu.tw}}}\inst{1}  \and 
        Tyng-Luh Liu\samethanks \inst{1,4}   \and
        Chi-Tai Fang\samethanks \inst{2,3} 
}
\authorrunning{Y-C Ho {\em et al}.}
%
\institute{Taiwan AI Labs, Taipei, Taiwan \\
\and Institute of Epidemiology and Preventive Medicine, College of Public Health, National Taiwan University, Taipei, Taiwan \\
\and Division of Infectious Diseases, Department of Internal Medicine, National Taiwan University Hospital, Taipei, Taiwan 
\and Institute of Information Science, Academia Sinica, Taiwan\\
}

\maketitle

\begin{abstract}
How to avoid a second wave of COVID-19 after reopening the economy is a pressing question. The extremely high basic reproductive number $R_0$ (5.7 to 6.4, shown in new studies) of SARS-CoV-2 further complicates the challenge. Here we assess effects of Social distancing 2.0, i.e. proximity alert (to maintain inter-personal distance) plus privacy-preserving contact tracing. To solve the dual task, we developed an open source mobile app. The app uses a Bluetooth-based, decentralized contact tracing platform over which the anonymous user ID cannot be linked by the government or a third party. Modelling results show that a 50\% adoption rate of Social distancing 2.0, with privacy-preserving contact tracing, would suffice to decrease the $R_0$ to less than 1 and prevent the resurgence of COVID-19 epidemic.
\end{abstract}

\section{Introduction}

The coronavirus disease 2019 (COVID-19) pandemic has caused 16,558,289 confirmed cases and 656,093 deaths worldwide by July 29, 2020 [1]. To limit the person-to-person transmission of its causative agent, SARS-CoV-2, governments imposed restrictive social distancing orders for mass quarantine at home (“lock-down”) [2, 3] along with rapid scaling-up of SARS-CoV-2 testing to detect and isolate infectious persons [4, 5]. The combined effect of these interventions successfully slowed the transmission of SARS-CoV-2 in South Korea, Europe, Japan, and the United States, with marked decrease in numbers of new cases and deaths [3, 5, 6]. However, modeling results predict that the COVID-19 epidemic will rebound after lifting of social distancing restriction [3]. States or countries which chose to end the lock-down early already experienced a resurgence in the numbers of new COVID-19 cases [6]. How to safely reopen the economy and simultaneously avoid a second wave of COVID-19 epidemic now becomes a pressing question for which there is still no clear answer.

Universal SARS-CoV-2 testing alone will not suffice to break the chains of ongoing transmission because SARS-CoV-2-infected persons start to be infectious 2–2.5 days before the onset of clinical symptoms [7-10]. To block this presymptomatic transmission, it is necessary to conduct contact tracing to inform those who had close contact with SARS-CoV-2-tested-positive persons (defined as: within 6 feet, for $\ge 15$ minutes, during the infectious period back to 48 hours before the onset of symptoms) to stay at home and avoid close contacts with others for 14 days [11, 12]. Since timeliness is of critical importance, instant digital contact tracing (through a contact-tracing app which records close contacts and notifies those people immediately upon a positive SARS-CoV-2 testing result) has been proposed as a powerful strategy to control the COVID-19 epidemic after the lift of lock-down [11].

Nevertheless, to date, privacy concerns continue to delay contact tracing apps in the United States [13]. In countries where COVID-19 contact tracing apps have been approved, the uptake remains low (25\% in Singapore, two months after launch, despite a target level of 75\%) [14, 15]. Surveys found that privacy concerns and the distrust towards governments are major reasons behind this low adoption rate [15]. The passive role of users (waiting a notification from someone unknown) is not appealing, either.

Moreover, to achieve sustained suppression of epidemic, interventions need to reduce the absolute value of basic reproductive number $R_0$ (the number of secondary cases generated by an infectious case in the absence of herd immunity) of SARS-CoV-2 to less than 1 [11]. Initial efforts to measure $R_0$ of SARS-CoV-2, based on epidemiological data in Wuhan, China, yielded estimates ranging from 2.2 to 2.7 [16, 17] which became the basis of previous modeling studies to assess the impact of interventions. However, new analyses of data across China found that SARS-CoV-2 is much more contagious than previously thought, with an $R_0$ as high as 5.7 [17]. This updated $R_0$ estimate is in line with the very high second attack rate (53.3\%–59.4\%) following exposure at community gatherings [18, 19] and the extremely fast growth rate of COVID-19 cases in the United States (from the initial imported cases in late January [20] to more than 1.8 million confirmed cases in late May 2020 [6] over a four-month period, with an $R_0$ estimate of 6.4, 95\% confidence interval: 6.0 to 6.8, see Fig. 1). Digital contact tracing to control epidemic under these new, much higher $R_0$ estimates has not been assessed.

Taiwan successfully controlled COVID-19 through early border control (14-day entry quarantine), moderate social distancing, SARS-CoV-2 testing, and contact tracing, with only 449 confirmed cases and 7 deaths by July 8, 2020 [21, 22]. Artificial intelligence-based information technology plays a central role behind the highly efficient implementation of these interventions [22-25]. Since early May 2020, the government has been contemplating how to reopen the society safely. Experts agreed that the privacy and data safety problems of using digital contact tracing must be solved. Furthermore, given the high contagiousness of SARS-CoV-2, digital contact tracing needs to be augmented by additional protective measures. 

We propose a new initiative, Social distancing 2.0, in which privacy-preserving contact tracing is supplemented with digital proximity alert to help people actively and voluntarily maintain safe inter-personal distance [26] after end of government-enforced compulsory social distancing orders (i.e. Social distancing 1.0). To solve the dual task, Taiwan AI Lab (https://covirus.cc/) developed an open source mobile app, Taiwan Social Distancing Application (TSD app) [27].

\begin{figure}[t!]
	\centering
	\includegraphics[width=0.92\textwidth]{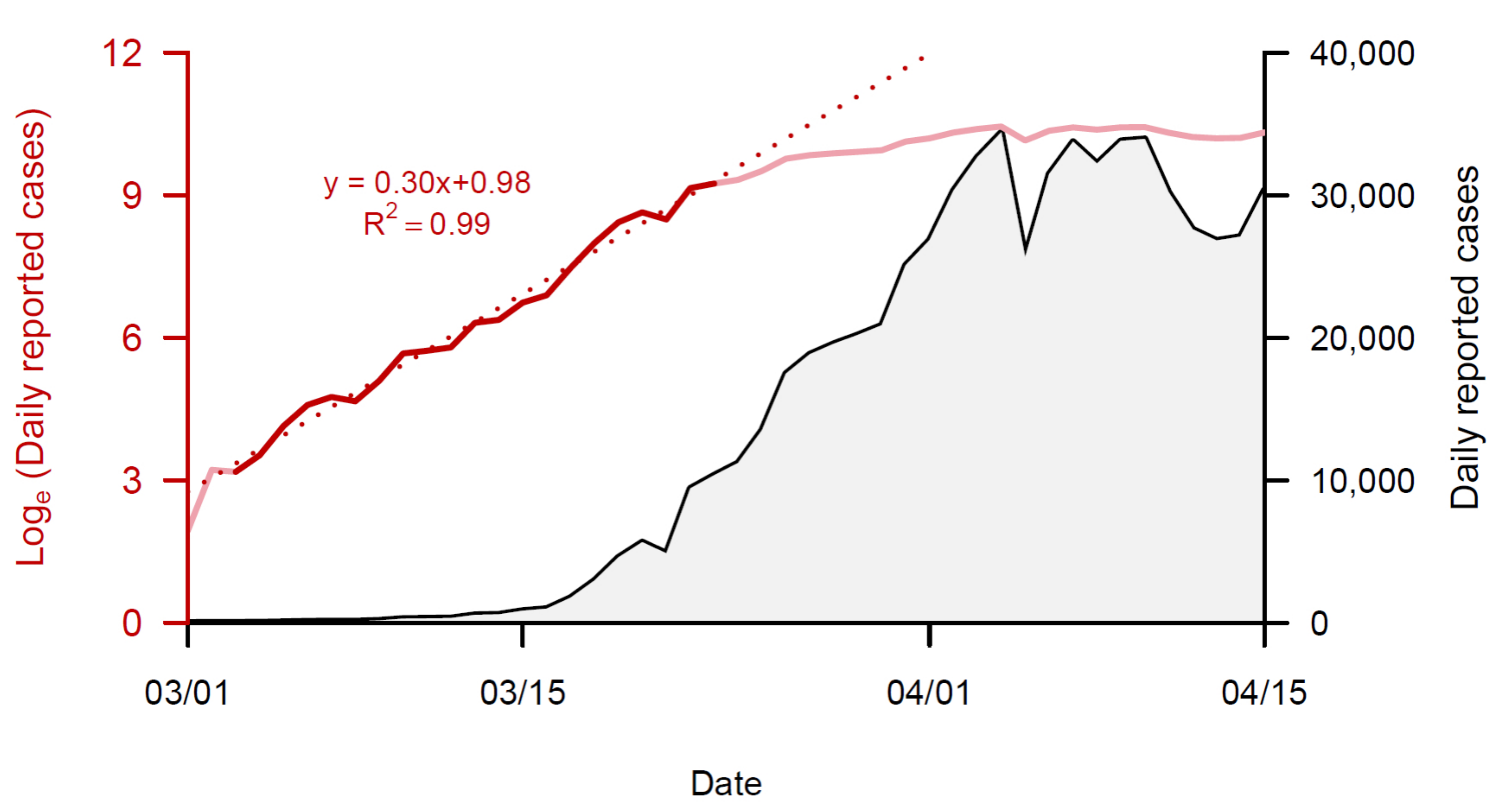}
	\caption{\textbf{ COVID-19 epidemic curve in the United States}. Data (6) are numbers of daily reported cases in log scale (curve in pink color) or linear scale (curve in black color). The epidemic curve followed an exponential growth from early to mid-March 2020 (curve in red color). The dot line (in red color) shows a linear regression over time (R2 = 0.99) for numbers of cases (in log scale) during this exponential phase, indicating an exponential growth rate of 0.30/day and a doubling time of 2.3 days. SIER model-fitting (see Supplementary material for details) yields an estimated $R_0$ of 6.4 (95\% confidence interval: 6.0 to 6.8).}
	\label{fig:epidemicCurve}
\end{figure}

\section{Taiwan Social Distancing App}
\label{sec:TSD}

Here we explain how privacy-preserving proximity alert and contact tracing is achieved in TSD app. We then evaluate the impact of Social distancing 2.0 on the $R_0$ and trajectory of COVID-19 epidemic.

\paragraph{Proximity Alert between TSD Users}
The TSD app, installed on mobile phones, communicates with other TSD clients by exchanging hashed ID via Bluetooth Low Energy (BLE) broadcast. The TSD app will calculate and record both distance and contact duration of nearby TSD clients. The app will send a proximity alert to remind the users to maintain safe inter-personal distance when another TSD client comes close to less than 6 feet for $\ge 15$ minutes.

\paragraph{Privacy-Preserving Contact Tracing}
At the core of the TSD app is a privacy-preserving contact tracing module [27]. The tracing method is designed based on analyzing the advantages and disadvantages of existing public algorithms [28-33], such as Decentralized Privacy-Preserving Proximity Tracing from EPFL [28], Privacy-Preserving Contact Tracing from Google and Apple [29], as well as BlueTrace from Singapore Government Technology Agency [30]. The TSD app meets the requirement of European Union General Data Protection Regulation (GDPR) [31]. Unlike other contact tracing apps, the TSD app uses a Bluetooth-based, decentralized contact tracing platform over which the anonymous user ID cannot be linked by the government or the third party. Each TSD app compares the public list of infected hashed IDs with its contact history on device and alerts the user if a high-risk contact history is detected.

\noindent {\em Data Policy}: We detail the data policy of TSD in four key aspects, including what data are collected, who can access these data, how the data are used, and when the life cycle of data ends. To protect the individual privacy, no personally identifiable information is collected by the app. TSD records only anonymous hashed IDs received, and stores their timestamp in the local device (the cell phone). The collected data, also known as the contact history, will never leave the device, and are kept for 28 days. When any of TSD users are confirmed as infected, they can provide public health organizations with the anonymous IDs used to broadcast themselves in the last 28 days. The public health organizations publish a list of IDs broadcast by known infected users, while the list is periodically updated at a defined time interval. These anonymous IDs cannot be linked back to the individual and will be destroyed after 7 days. The TSD app routinely checks the published list of infected anonymous IDs to match its local contact history and alerts its user when a high-risk contact history is found.

\paragraph{Contact Tracing Algorithm}
Driven by privacy preserving, the algorithm is designed based on the following five principles: 1) Notify people at risk of infection and give guidance on the next steps efficiently; 2) Minimize the amount of data collected in the process; 3) Prevent abuse of data thoroughly; 4) Avoid central tracking of non-infected users; and  5) Dismantle the app automatically after the outbreak is over.

{\em Registration Free}: The algorithm requires no registration step before using TSD to trace user’s contact history. All the collected data are generated by the app itself and cannot be linked back to the user. However, additional personal information might be needed when a user is contacted by public health organizations under the condition of infection or other emergencies. 
Anonymous IDs: Since the algorithm relies on the BLE protocol, the length of an anonymous ID is thus limited by the protocol’s payload size and should not overburden the local storage of the device. The length of a generated anonymous key is 128 bits. To avoid location tracking via exploring broadcasted IDs, the app must frequently change its anonymous ID. Anonymous IDs expire in 15 minutes, and at most 96 unique, secure, anonymous IDs will be generated each day.  The short rotation interval is adopted to prevent large-scale location tracking and avoid linking anonymous IDs back to an individual based on other real-world information. The TSD app stores all IDs it generates in the local device storage; however, IDs older than 28 days will be discarded for personal data protection and keep storage usage at a reasonable level.

{\em Contact History}: More crucially, the TSD app also locally records each ID it received via BLE broadcast. A timestamp is also logged. Each contact record requires 46 bytes (44 bytes hashed base64 string ID, 1 byte timestamp and 1 byte RSSI). Assuming that 10k IDs are received daily, about 12.28MB is required to save 28 days’ worth of data for contact history.

{\em Report when Infected}: When TSD users are tested positive, they first hash the IDs they have broadcast in the past 28 days using a crypto-secure hash function (e.g., SHA256). After providing basic contact information to a public health organization, they can upload the hashed anonymous IDs with an authorization code to the backend server. The server aggregates all newly infected hashed IDs, removes outdated IDs, and then publishes a new alert list to be downloaded. The backend is simple storage for infected IDs with no additional processing. For efficiency, the list can be shared via CDN or distributed servers, which lower the bandwidth cost of the backend. The backend server will also remove infected IDs older than 7 days to reduce resource usage and prevent abusive use of data, while still being valuable about containing the outbreak. To further anonymize the data when the infection case number is small, we continue to explore potential solutions such as adding spurious tokens or introducing mixing servers. 

{\em Risk Notification}: Each 4 hours, the TSD app will download an updated list of infected hashed IDs from the server and detect matches between the list and the local contact history. If a match is found, the user’s risk score is calculated based on estimated distance and contact duration. It will notify the user and provide health care information when the calculated risk score is high.

\paragraph{Security and Privacy}
Like other software systems, the TSD app could face a variety of adversarial threats and security concern is a top priority in designing the algorithm. Due to the nature of the BLE broadcast, adversaries can always record a received message and replay it back to the public at different locations before the ID is replaced. If the owner of the ID is diagnosed as infected, the replayed message will create multiple false positives to other TSD users. However, such an online attack demands large scale of device deployment to record infected IDs before they are published, and thus requires extensive budget which is not feasible to small-scale adversaries. The security of TSD can also be challenged from perverse infected users by hiding their contact history via removing the app, refusing to share their past IDs, disabling BLE broadcast whenever they want. Finally, adversarial attacks such as jamming attack could cause deny of service for the BLE-based app. To strengthen TSD against the various security attacks is still an ongoing research effort, and the progress will be constantly updated.

\begin{figure}[t]
	\centering
	\includegraphics[width=0.80\textwidth]{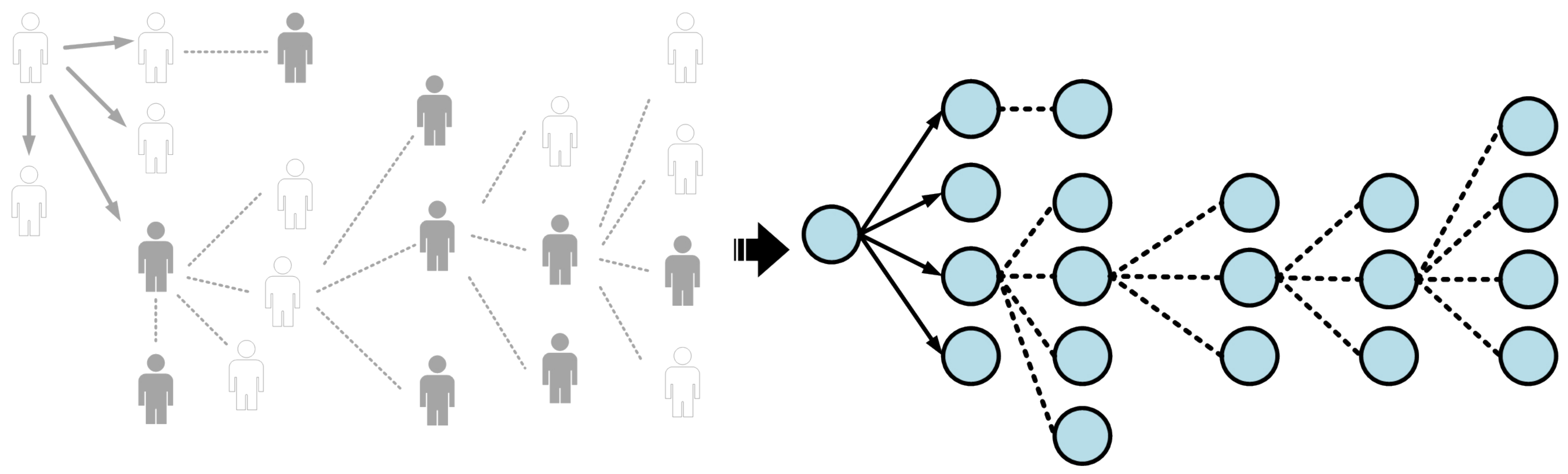}
	\caption{\textbf{Global interaction graph}. The social relationships among all users.}
	\label{fig:globalG}
\end{figure}

Privacy preserving can be better understood through two graph models accounting for global interaction and proximity information. The global interaction graph (Fig. 2) reflects the social relationships of all users in the TSD system. The interaction between two users is revealed only to themselves. Infected users will not share their contact history with the authority but the past anonymous IDs to the public health organizations. Under such a design, no party can learn the global interaction graph from the system. The proximity graph (Fig. 3) encodes contacts between infected users and other individuals, which is essential to the contact tracing application. However, only a relevant subset (i.e., related to infected users) of the proximity graph is revealed to each user. Users will not share their subgraph with other users or authorities. Notice that the anonymous IDs in our design are unlinkable. Only the device that generated these keys knows their relationship. When a user is confirmed infected, the anonymous IDs they have used in the last 28 days will be shared with the public. The user’s device will rotate to a new set of IDs after the user is diagnosed as infected. Hence, those published infected IDs cannot be used to link back to the user. For users other than infected ones, their data will not leave their device at all, and thus no information about them is leaked to any party.

\begin{figure}[!h]
	\centering 
	\hspace{1.50cm}
	\includegraphics[width=0.80\textwidth]{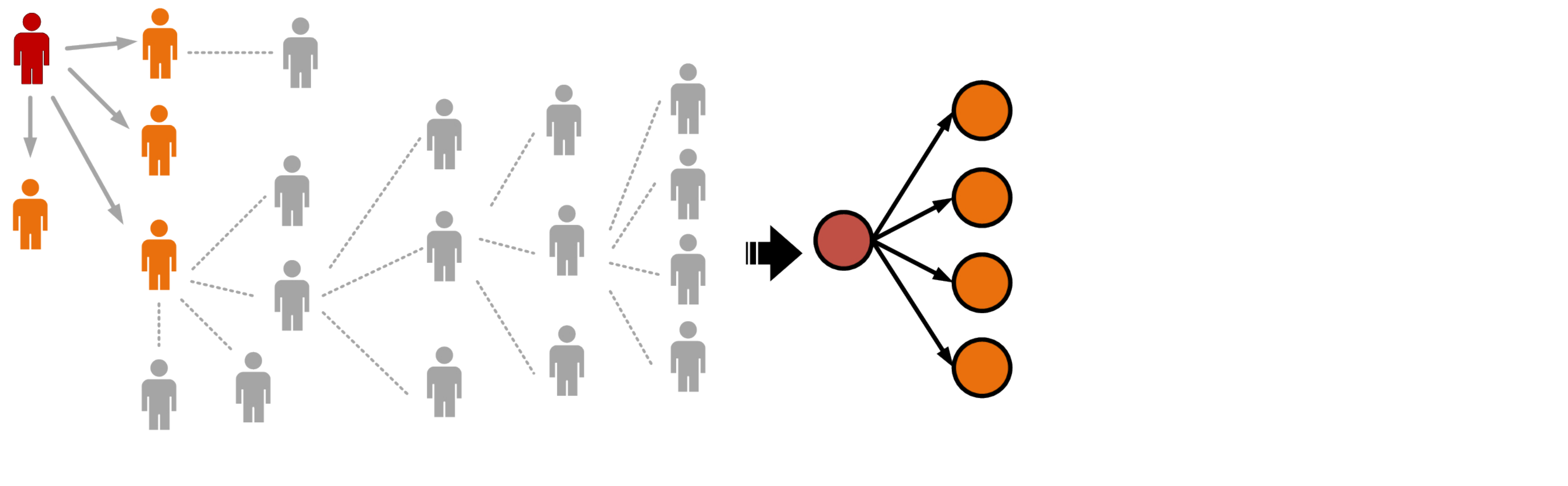}
	\caption{\textbf{Proximity graph}. Those who (colored as orange) are within the concerned contact range of an anonymous infected user (colored as red).}
	\label{fig:proximityG}
\end{figure}

\section{Social Distancing 2.0 and Its Impacts}
\label{sec:impacts}

\paragraph{Impact of Social Distancing 2.0 on $R_0$}
We applied an SEIR model to simulate the effect of Social distancing 2.0 on $R_0$ of SARS-CoV-2 (see Supplementary Material). The model followed the general analytical framework of Grad and Lipsitch et al. [34] as well as Giordano et al. [5]. New empirical studies now provided more accurate estimates on key parameters, including basic reproductive number before interventions [17], pre-symptomatic infectious period [8], and infectious duration after the onset of illnesses [9]. New systematic reviews and meta-analyses provided precise estimates on the protective efficacy of maintaining $a >6$ feet ($\sim\,$1.8 meter) inter-personal distance against SARS-CoV-2 transmission [26]. (Table 1).
$R_0$ is estimated by the following formula: (see Supplementary Material for details)

\begin{equation}
\mathrm{Total}\; R_0 = \frac{\beta(1-f p)\cdot\frac{\sigma(1-q_A)}{\mu+\sigma}\cdot\left[1-qc\cdot \left(\frac{d+\gamma \cdot q_H}{\mu+\gamma + d}\right)\right]}{\mu+\gamma+d}+\frac{\beta(1-f p)\cdot\frac{\sigma  q_A}{\mu+\sigma}}{\mu+\gamma}
\label{eqn:r0}
\end{equation}
\noindent The first term in (\ref{eqn:r0}) is the contribution from symptomatic patients (including transmissions occurred in the pre-symptomatic infectious period), while the second term is the contribution from asymptomatic infected persons. We note here that the various symbols in (\ref{eqn:r0}) respectively represent:
\begin{itemize}
\item $\beta$ is the transmission coefficient. 
\item $f$ is level of social restriction (Social distancing 1.0) or adoption rate (Social distancing 2.0 apps).
\item $p$ is compliance (Social distancing 1.0) or protective efficacy of distancing (Social distancing 2.0).
\item $1/\sigma$ is latency period (before the start of pre-symptomatic infectious period).
\item $q_A$ is the proportion of infected persons who remain asymptomatic throughout the course.
\item $1/\mu$ is life expectancy in the United States.
\item $qc$ is adoption rate of contact tracing app (without proximity alert) or Social distancing 2.0 app (with dual-function of proximity alert and privacy-preserving contact tracing).
\item $1/d$ is mean duration from symptom onset to isolation.
\item $1/\gamma$ is mean duration of infectiousness.
\item $q_H$ is proportion of symptomatic patients who developed several illnesses and need hospitalization.
\end{itemize}

\begin{figure}[ht!]
    \setlength{\abovecaptionskip}{-65pt} 
	\centering
	\includegraphics[width=0.98\textwidth]{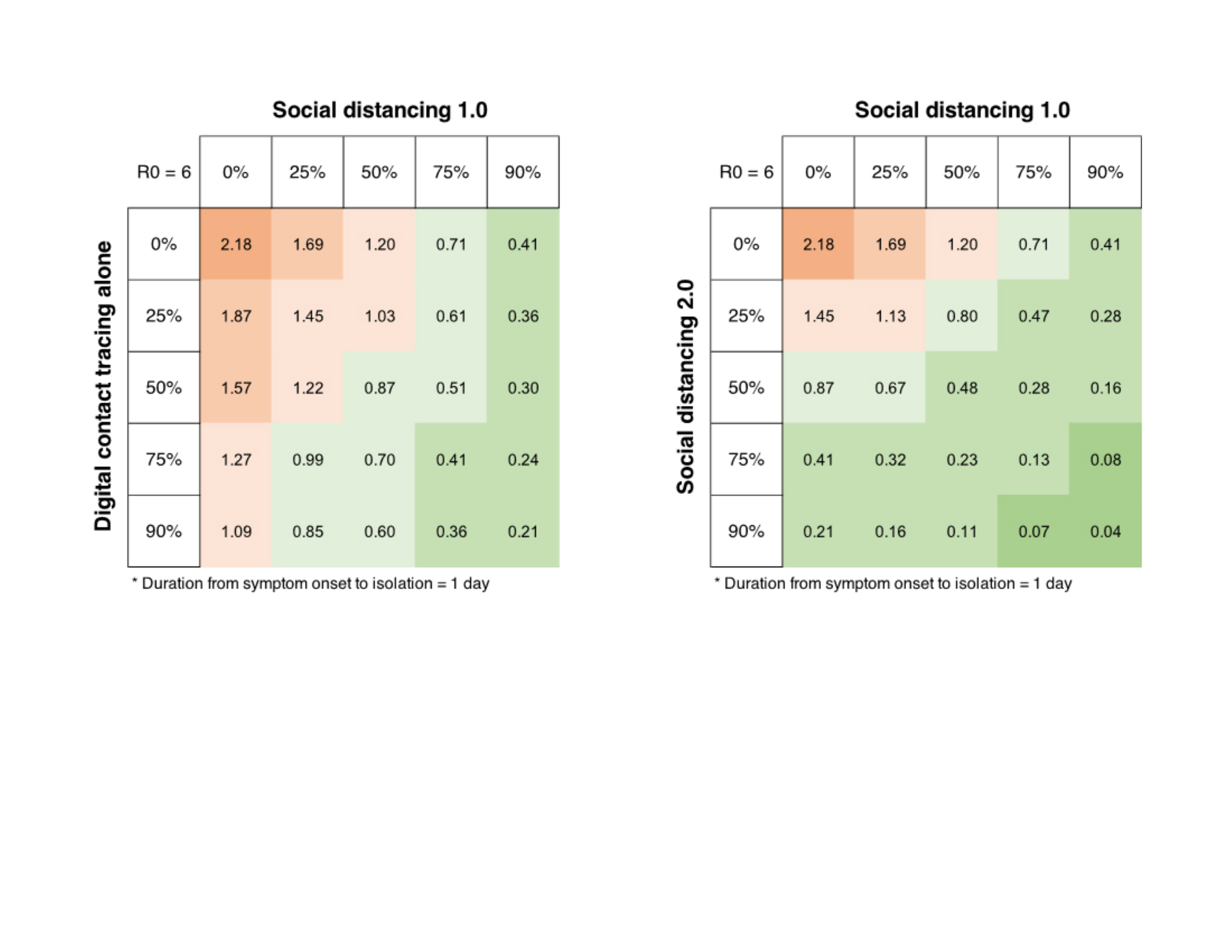}
	\caption{\textbf{Quantifying $R_0$ under Social distancing 2.0}. Simulations are conducted with a baseline $R_0$ at the level of 6.0 (Fig.~\ref{fig:epidemicCurve}) (17), under a setting with universal rapid SARS-CoV-2 testing that ensures rapid diagnosis and isolation of symptomatic patients (mean duration from the symptoms onset to isolation: 1 day). Columns show levels of compulsory Social distancing 1.0 (\% decrease in social contacts). Rows show levels of uptake for Social distancing 2.0 apps (\% adoption rates). Rapid isolation of symptomatic patients alone decreases $R_0$ from 6 to 2.18. The $R_0$ can be further reduced to 0.71, with epidemic control, by restricting 75\% social contacts under Social distancing 1.0. Nevertheless, $R_0$ will easily rebound to values higher than 1 after easing social distancing 1.0 restriction to 50\% or lower. Left panel: Effect of digital contact tracing alone (without proximity alert to keep a $ > 6$~feet inter-personal distance). When baseline $R_0$ is at the level of 6.0, even with a 90\% uptake rate, contact tracing apps alone will not bring the $R_0$ back to below 1 after the end of Social distancing 1.0. Right panel: Effect of Social distancing 2.0 (digital contact tracing plus proximity alert to keep a $> 6$ feet inter-personal distance). A 50\% adoption rate of such apps would suffice to decrease $R_0$ to less than 1, with sustained suppression of epidemic, after the end of Social distancing 1.0.}
	\label{fig:R0}
\end{figure}

Figure~\ref{fig:R0} shows that, when baseline $R_0$ is at the level of 6.0, even a 90\% uptake rate of single function contact tracing apps (without proximity alert) will not bring the $R_0$ back to below 1 after lift-off of lock-down. In contrast, a 50\% adoption rate of dual function Social distancing 2.0 app (contact tracing plus proximity alert) would suffice to decrease $R_0$ to less than 1 and ensure the sustained suppression of COVID-19 after reopening

\begin{figure}[ht!]
	\centering
	\includegraphics[width=0.95\textwidth]{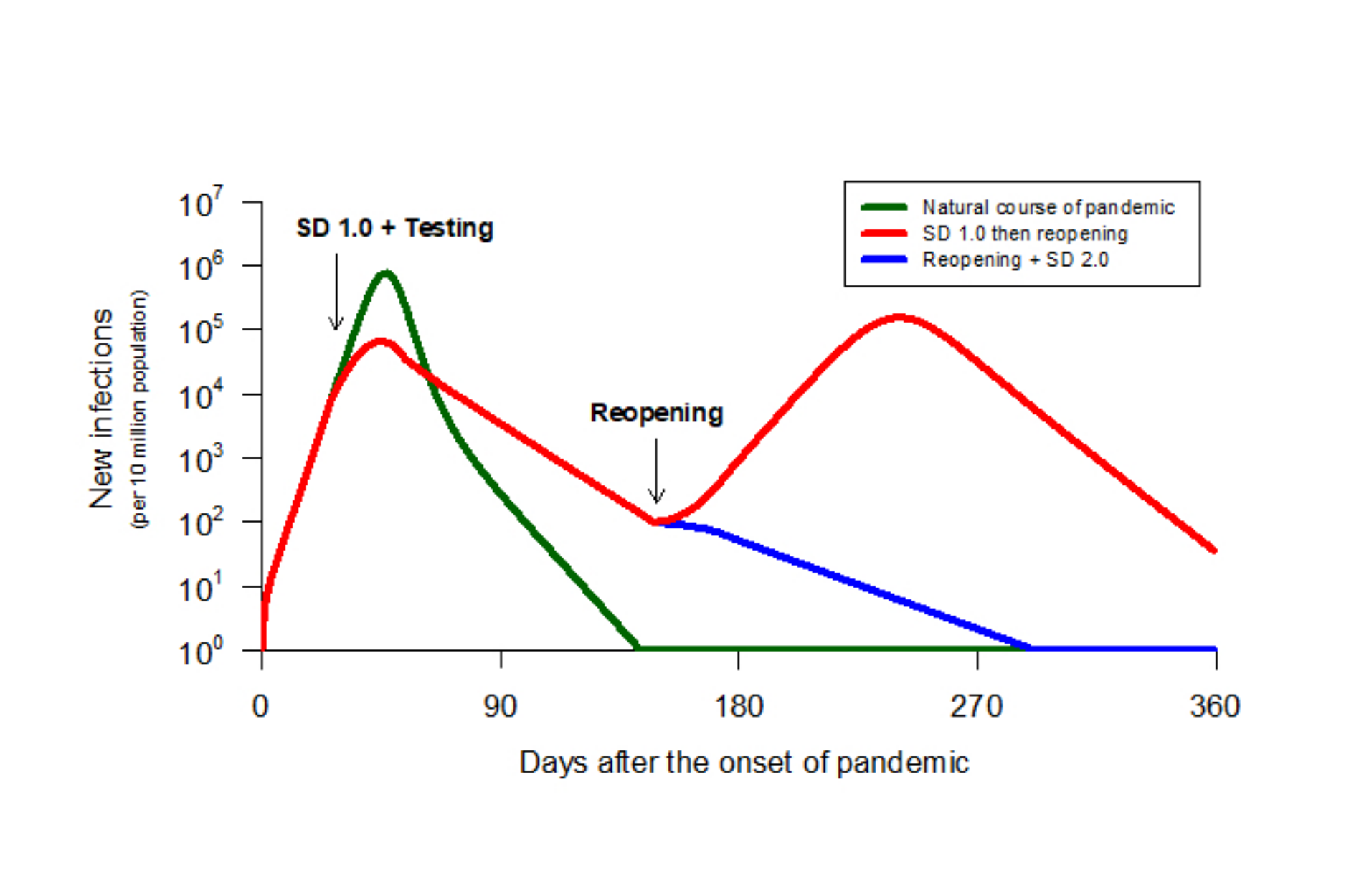}
	\caption{\textbf{Trajectory of COVID-19 epidemic without or with Social distancing 2.0}. Simulations are conducted with a baseline $R_0$ at the level of 6.0 (Fig.~\ref{fig:epidemicCurve}) (17), in a 10-million population area. The green line shows numbers of daily new infections when the pandemic runs its nature course. Compulsory social distancing (SD 1.0) (lock-down) to restrict social contacts by 75\% started on day 27 (marked by arrow), along with universal rapid SARS-CoV-2 testing for prompt isolation of symptomatic patients to achieve a final mean duration of 1 day from symptom onset to isolation, (both interventions are rolled out over a 4-week period) flattens the epidemic curve. The red line shows that SD 1.0 plus testing decreases the peak incidence of new infections by more than 90\%. COVID-19 epidemic rebounds after gradually reopening started from day 148 (marked by arrow) over a 4-weeks period, causing a second wave of COVID-19 epidemic with a magnitude larger than that of the first wave (red line, right half). The launch and rollout of Social distancing 2.0 app (SD 2.0) to an adoption rate of 50\% over a 4-week period after the start of re-opening, nevertheless, prevents the second wave of COVID-19 pandemic (blue line). Moreover, SD 2.0 will continue to suppress COVID-19 until its elimination (when numbers of daily new infection becomes less than 1). Sensitivity analysis shows that moderate variations in SD 2.0 apps adoption rate (from 45\% to 55\%) does not alter the protective effect of SD 2.0 against a resurgence of COVID-19 pandemic (blue shadow area).}
	\label{fig:Trajectory}
\end{figure}

\vspace{5cm}

\paragraph{Impact of Social Distancing 2.0 on Trajectory of Epidemic}
Figure~\ref{fig:Trajectory} shows that, consistent with previous studies [3, 34], COVID-19 epidemic will rebound after reopening. The second wave of epidemic will have a magnitude even greater than that of the first wave. The launch and rollout of Social distancing 2.0 app to an adoption rate of 50\% after re-opening prevents the second wave of COVID-19 pandemic. Sensitivity analysis shows that moderate variations in Social distancing 2.0 apps adoption rate (from 45\% to 55\%) does not alter its protective effect against a resurgence of COVID-19.

\section{Discussions}
\label{sec:discussion}

Our Social distancing 2.0 initiative simultaneously addresses two unmet and seemingly conflicting societal needs: first, avoiding a second wave of COVID-19 after reopening the economy; second, upholding privacy, data security, and democratic values in a time of pandemic [35]. Proximity alert function in Social distancing 2.0 apps turns user experience from passively waiting bad news into actively protecting oneself and others, while privacy-preserving contact tracing prevents abuse of digital tracing by authoritarian regimes or malicious hackers. Dual function Social distancing 2.0 apps effectively suppress the COVID-19 epidemic at 50\% adoption rate, in contrast to failure to suppress $R_0$ to below 1 by single function contact tracing app even when its uptake rate reaches 90

Social distancing is a traditional standard practice to curb the outbreak of an infectious pandemic. However, it causes extreme disruption of society and economy, and therefore, is not sustainable for period long enough to achieve a durable suppression. The Social distancing 2.0 initiative is a smart and non-disruptive continuing implementation of de facto social distancing made possible by the cutting-edge modern technology, which allows nearly normal social and economic activities to proceed safely. This will not only avoid a second wave of COVID-19 epidemic in late 2020 and beyond, but also save millions of lives before the universal availability of medical masks and effective vaccines.

\begin{table}
  \caption{Key parameters of the SEIR model for SARS-CoV-2 transmission and pathogenesis}
  \label{tbl:parameters}
  \centering
  \begin{tabular}{| p{6cm} | c | p{4cm} |}
    \hline
    \multirow{2}{*}{{\bf Parameters}}     & \multirow{2}{*}{{\bf Value (mean)}} & \multirow{2}{*}{{\bf References}}  \\
    & &  \\
    \hline 
    Pre-intervention basic reproductive number $R_0$ of SARS-CoV-2 
    &
    \raisebox{-1.5ex}{6.0}
    &
    5.7 [17] $\qquad$ \hspace{2.22cm} \linebreak 6.4 (Fig.~\ref{fig:epidemicCurve}, this study) \\
    \hline
    Incubation period (from infection to the onset of illnesses)
    &
    \raisebox{-1.7ex}{5.2 days}
    & 
    \raisebox{-1.7ex}{[16]} \\
    \hline
    Pre-symptomatic infectious period before the onset of illnesses
    &
    \raisebox{-1.7ex}{2.5 days}
    & 
    \raisebox{-1.7ex}{[8]} \\
    \hline
    \raisebox{-5.8ex}{Latent period (from infection to the onset} of infectiousness)
    &
    \raisebox{-5.8ex}{2.7 days} 
    &
    5.2 days (incubation period) [16] minus 2.5 days (pre-symptomatic infectious period before the onset of illnesses) [8, 9] \\
    \hline
    Infectious period after the onset of illnesses
    &
    6.0 days
    &
    [9] \\
    \hline
    Proportion of severe/critical cases among all infected persons 
    &
    \raisebox{-1.5ex}{19\%}
    &
    \raisebox{-1.5ex}{[36]} \\
    \hline
    Infection-related mortality among sever/critical cases 
    &
    \raisebox{-1.5ex}{12\%}
    & 
    \raisebox{-1.5ex}{[36]} \\
    \hline
    Proportion of infected persons who remain asymptomatic throughout course 
    &
    \raisebox{-1.5ex}{10\%}
    &
    \raisebox{-1.5ex}{[7,37]} \\
    \hline
    Protective efficacy of maintaining a less than 15 minutes and $>$ 6 feet physical distance during contacts, in terms of reduction in probability of transmission per contact 
    &
    \raisebox{-4.8ex}{90\%}
    &
    \raisebox{-4.8ex}{[26]} \\
    \hline
  \end{tabular}
\end{table}

\section*{Acknowledgments}
We thank Chi-Mai Chen, Vice Premier of Taiwan, for initiating this cross-disciplinary project. Funding: Ministry of Health and Welfare and National Taiwan University Infectious Diseases Research and Education Center, Taipei, Taiwan (grant no. MOHW-109-CDC-ZH108028). Author contributions: Conceptualization: YCT, TLL, YHC, CTF. Data curation: YHC. Funding acquisition: YCT, CTF. Software: YCH, SHH, CHH, PP, CHC, DKY, YCT, TLL. Methodology: CTF. Modelling analysis: YHC, CTF. Writing, original draft: CHC, YHC, CTF. Writing, review and editing: all authors. Competing interests: None declared. Data and materials availability: All data is available in the main text or the supplementary materials. The source code of TSD app is openly available at Taiwan AI Lab website: https://covirus.cc/.

\section*{References}

\small

[1] World Health Organization: Coronavirus disease 2019 (COVID-19) Situation Report 191 (July 29, 2020). \url{https://www.who.int/docs/default-source/coronaviruse/situation-reports/20200729-covid-19-sitrep-191.pdf?sfvrsn=2c327e9e_2}.

\vspace{0.1cm}
\noindent
[2]	Proclamation by the governor: amending proclamations 20–05; 20–25, stay home–stay healthy. Washington, USA. 2020 Mar 23 [cited June 3, 2020]. \url{https://reurl.cc/ZO6eKp}. 

\vspace{0.1cm}
\noindent
[3]	L. Matrajt, T. Leung, Evaluating the effectiveness of social distancing interventions to delay or flatten the epidemic curve of coronavirus disease. Emerg Infect Dis 26, [cited June 3, 2020]. \url{https://doi.org/2010.3201/eid2608.201093} (2020).

\vspace{0.1cm}
\noindent
[4]	A. V. Dora et al., Universal and serial laboratory testing for SARS-CoV-2 at a long-term care skilled nursing facility for veterans - Los Angeles, California, 2020. MMWR Morb Mortal Wkly Rep 69, 651-655 (2020).

\vspace{0.1cm}
\noindent
[5]	G. Giordano et al., Modelling the COVID-19 epidemic and implementation of population-wide interventions in Italy. Nat Med, 1-6 (2020).

\vspace{0.1cm}
\noindent
[6]	COVID-19 coronavirus pandemic: country-specific data on new cases and new death. \url{https://www.worldometers.info/coronavirus/} (accessed on June 3, 2020).

\vspace{0.1cm}
\noindent
[7]	M. M. Arons et al., Presymptomatic SARS-CoV-2 Infections and Transmission in a Skilled Nursing Facility. N Engl J Med 382, 2081-2090 (2020).

\vspace{0.1cm}
\noindent
[8]	X. He et al., Temporal dynamics in viral shedding and transmissibility of COVID-19. Nat Med 26, 672-675 (2020).

\vspace{0.1cm}
\noindent
[9]	H. Y. Cheng et al., Contact tracing assessment of COVID-19 transmission dynamics in Taiwan and risk at different exposure periods before and after symptom onset. JAMA Intern Med, (2020).

\vspace{0.1cm}
\noindent
[10]	H. Nishiura, N. M. Linton, A. R. Akhmetzhanov, Serial interval of novel coronavirus (COVID-19) infections. Int J Infect Dis 93, 284-286 (2020).

\vspace{0.1cm}
\noindent
[11]	L. Ferretti et al., Quantifying SARS-CoV-2 transmission suggests epidemic control with digital contact tracing. Science 368, (2020).

\vspace{0.1cm}
\noindent
[12]	Centers for Disease Control and Prevention. Case investigation and contact tracing : part of a multipronged approach to fight the COVID-19 pandemic. [cited June 3, 2020]. \url{https://www.cdc.gov/coronavirus/2019-ncov/php/principles-contact-tracing.html}.

\vspace{0.1cm}
\noindent
[13]	O. L. Gallaga, Privacy concerns continue to delay contact tracing apps. [cited on June 3, 2020] \url{https://www.governing.com/security/Privacy-Concerns-Continue-to-Delay-Contact-Tracing-Apps.html}.

\vspace{0.1cm}
\noindent
[14]	Stephanie Findlay in New Delhi, Stefania Palma in Singapore and Richard Milne in Oslo. May 18, 2020. Financial Times: Coronavirus contact-tracing apps struggle to make an impact [cited June 4, 2020] \url{https://www.ft.com/content/21e438a6-32f2-43b9-b843-61b819a427aa}.

\vspace{0.1cm}
\noindent
[15]	Dewey Sim in Singapore and Kimberly Lim. May 18, 2020. South China Morning Post: Why aren’t Singapore residents using the TraceTogether contact-tracing app? [cited on June 4, 2020] \url{https://www.scmp.com/week-asia/people/article/3084903/coronavirus-why-arent-singapore-residents-using-tracetogether}.

\vspace{0.1cm}
\noindent
[16]	Q. Li et al., Early Transmission Dynamics in Wuhan, China, of Novel Coronavirus-Infected Pneumonia. N Engl J Med 382, 1199-1207 (2020).

\vspace{0.1cm}
\noindent
[17]	S. Sanche et al., High Contagiousness and Rapid Spread of Severe Acute Respiratory Syndrome Coronavirus 2. Emerg Infect Dis 26. \url{https://doi.org/10.3201/eid2607.200282} (2020).

\vspace{0.1cm}
\noindent
[18]	L. Hamner et al., High SARS-CoV-2 Attack Rate Following Exposure at a Choir Practice - Skagit County, Washington, March 2020. MMWR Morb Mortal Wkly Rep 69, 606-610 (2020).

\vspace{0.1cm}
\noindent
[19]	A. James et al., High COVID-19 Attack Rate Among Attendees at Events at a Church - Arkansas, March 2020. MMWR Morb Mortal Wkly Rep 69, 632-635 (2020).

\vspace{0.1cm}
\noindent
[20]	CDC COVID-19 Response Team, Evidence for limited early spread of COVID-19 within the United States, January–February 2020. MMWR ePub: 29 May 2020. DOI: http://dx.doi.org/10.15585/mmwr.mm6922e1external, (2020).

\vspace{0.1cm}
\noindent
[21]	Taiwan Centers for Disease Control: COVID-19 statistics. [cited on July 8, 2020] \url{https://www.cdc.gov.tw/En}.

\vspace{0.1cm}
\noindent
[22]	C. J. Wang, C. Y. Ng, R. H. Brook, Response to COVID-19 in Taiwan: big data analytics, new technology, and proactive testing. JAMA, (2020).

\vspace{0.1cm}
\noindent
[23]	C. M. Chen et al., Containing COVID-19 among 627,386 persons in contact with the Diamond Princess cruise ship passengers who disembarked in Taiwan: big data analytics. J Med Internet Res 22, e19540 (2020).

\vspace{0.1cm}
\noindent
[24]	Taiwan AI Lab: Taiwan set up AI tools tracing virus transmission with sequencing and multi-omics data. [cited June 4, 2020] \url{https://covirus.cc/phylogeny-intro.html} (2020).

\vspace{0.1cm}
\noindent
[25]	C. F. Yeh et al., A cascaded learning strategy for robust COVID-19 pneumonia chest X-ray screening. [cited June 4, 2020] \url{https://arxiv.org/abs/2004.12786} (2020).

\vspace{0.1cm}
\noindent
[26]	D. K. Chu et al., Physical distancing, face masks, and eye protection to prevent person-to-person transmission of SARS-CoV-2 and COVID-19: a systematic review and meta-analysis. [cited on June 7, 2020] DOI: 10.1016/s0140-6736(20)31142-9. Lancet, (2020).

\vspace{0.1cm}
\noindent
[27]	Taiwan AI Lab: Taiwan Social Distancing Application (TSD app). [cited on June 5, 2020) \url{https://covirus.cc/social-distancing-app-intro.html} (2020).

\vspace{0.1cm}
\noindent
[28]	DP3T: Decentralized Privacy-Preserving Proximity Tracing. [cited on June 4, 2020] \url{https://github.com/DP-3T/documents} (2020).

\vspace{0.1cm}
\noindent
[29]	Privacy-Preserving Contact Tracing by Apple/Google. [cited on June 5, 2020] \url{https://www.apple.com/covid19/contacttracing/} (2020).

\vspace{0.1cm}
\noindent
[30]	BlueTrace: a privacy-preserving protocol for community-driven contact tracing across border. [cited on June 5, 2020] \url{https://bluetrace.io/} (2020).

\vspace{0.1cm}
\noindent
[31]	Commission Recommendation (EU) 2020/518 of 8 April 2020 on a common Union toolbox for the use of technology and data to combat and exit from the COVID-19 crisis, in particular concerning mobile applications and the use of anonymised mobility data. [cited on June 6, 2020] \url{https://eur-lex.europa.eu/legal-content/EN/TXT/?uri=CELEX:32020H0518} (2020).

\vspace{0.1cm}
\noindent
[32]	H. Cho, D. Ippolito, Y. W. Yu, Contact tracing mobile apps for COVID-19: privacy considerations and related trade-offs. [cited on June 5, 2020] \url{https://arxiv.org/abs/2003.11511}.

\vspace{0.1cm}
\noindent
[33]	J. Stanley, J. S. Granick, ACLU white paper: The limits of location tracking in an epidemic [April 8, 2020]. [cited on June 5, 2020] \url{https://www.aclu.org/report/aclu-white-paper-limits-location-tracking-epidemic} (2020).

\vspace{0.1cm}
\noindent
[34]	S. M. Kissler, C. Tedijanto, E. Goldstein, Y. H. Grad, M. Lipsitch, Projecting the transmission dynamics of SARS-CoV-2 through the postpandemic period. Science 368, 860-868 (2020).

\vspace{0.1cm}
\noindent
[35]	M. M. Mello, C. J. Wang, Ethics and governance for digital disease surveillance. Science 368, 951-954 (2020).

\vspace{0.1cm}
\noindent
[36]	The Novel Coronavirus Pneumonia Emergency Response Epidemiology Team., The Epidemiological Characteristics of an Outbreak of 2019 Novel Coronavirus Diseases (COVID-19). China CDC Weekly 2, 113-122 (2020).

\vspace{0.1cm}
\noindent
[37]	S. C. Chang, Proportion of asymptomatic SARS-CoV-2 infections among 398 laboratory-confirmed SARS-CoV-2-infected patients in Taiwan [in traditional Chinese]. April 18, 2020 [cited June 7, 2020]. \url{https://news.ltn.com.tw/news/life/breakingnews/3137967} (2020).

\vspace{0.1cm}
\noindent
[38] 	J. Wallinga, P. Teunis, M. Kretzschmar, Using data on social contacts to estimate age-specific transmission parameters for respiratory-spread infectious agents. Am J Epidemiol 164, 936-944 (2006)

\vspace{0.1cm}
\noindent
[39]	M. J. Keeling, P. Rohani, Modeling Infectious Diseases in Humans and Animals.  Princeton University Press. (2008)

\vspace{0.1cm}
\noindent
[40]	The World Bank. Life expectancy at birth - United States. \url{https://data.worldbank.org/indicator/SP.DYN.LE00.IN?locations=US} (2019).

\vspace{0.1cm}
\noindent
[41] 	C. Huang, Y. Wang, X. Li, L. Ren, J. Zhao, Y. Hu, L. Zhang, G. Fan, J. Xu, X. Gu, Z. Cheng, T. Yu, J. Xia, Y. Wei, W. Wu, X. Xie, W. Yin, H. Li, M. Liu, Y. Xiao, H. Gao, L. Guo, J. Xie, G. Wang, R. Jiang, Z. Gao, Q. Jin, J. Wang, B. Cao, Clinical features of patients infected with 2019 novel coronavirus in Wuhan, China. Lancet 395, 497-506 (2020).

\end{document}